\documentclass[aps,twocolumn,showkeywords,floatfix,superscriptaddress]{revtex4-1}
\usepackage[linktocpage,bookmarksopen,bookmarksnumbered]{hyperref}
\usepackage[utf8]{inputenc} 
\usepackage{graphicx}
\usepackage{diagbox}
\usepackage{amsfonts}
\usepackage{amsmath} 
\usepackage[usenames]{color}

\begin{document}

\title{Analytic continuation via domain-knowledge free machine learning}

\author{Hongkee Yoon}

\affiliation{Department of Physics, KAIST, 291 Daehak-ro, Yuseong-gu, Daejeon
	34141, Republic of Korea }

\author{Jae-Hoon Sim}

\affiliation{Department of Physics, KAIST, 291 Daehak-ro, Yuseong-gu, Daejeon
	34141, Republic of Korea }

\author{Myung Joon Han}
\email{mj.han@kaist.ac.kr}

\affiliation{Department of Physics, KAIST, 291 Daehak-ro, Yuseong-gu, Daejeon
	34141, Republic of Korea }

\begin{abstract}
	We present a machine-learning approach to 
	a  long-standing issue in quantum many-body physics, namely, 
	analytic continuation.
	This notorious ill-conditioned problem of obtaining spectral function
	from imaginary time Green's function has been a focus of
	new method developments for past decades.
	Here we demonstrate the usefulness of modern machine-learning techniques
	including convolutional neural networks and the variants of stochastic 
	gradient descent  optimiser.
	Machine-learning continuation kernel is successfully realized
	without any `domain-knowledge', which means that any physical `prior' is not utilized in the kernel construction and the neural networks `learn' the knowledge solely from `training'.
	The outstanding performance is achieved
	for both insulating and metallic band structure.
	Our machine-learning-based approach not only provides the more accurate
	spectrum than the conventional methods
	in terms of peak positions and heights,
	but is also more robust against the noise 
	which is the required key feature for any 
	continuation technique to be successful.
	Furthermore, its computation speed
	is 10$^4$--10$^5$ times faster than maximum entropy method.
\end{abstract}

\date{\today}

\maketitle

\section{Introduction}

Matsubara Green's function method is a useful theoretical tool for 
quantum many-body problems. While the calculation often becomes much
more tractable in the imaginary time (or equivalently, frequency)
domain, working with Matsubara function inevitably introduces other
theoretical difficulties. One of the most typical cases happens when
one tries to obtain a spectral function (or any other measurable
quantity), which is defined in real frequency space, from imaginary Green's
function. This procedure is known as `analytic continuation' and
poses a notorious ill-conditioned inverse problem. The severe
noise sensitivity significantly undermines the predictability and
the usefulness of theoretical methods such as quantum Monte Carlo
(QMC). Many different approaches have been suggested to solve this
problem including Pade approximation 
\cite{vidberg_solving_1977,gunnarsson_analytical_2010-1},
maximum entropy (MEM)
\cite{haule_dynamical_2010,bergeron_algorithms_2016,jarrell_bayesian_1996,gunnarsson_analytical_2010},
and stochastic method \cite{sandvik_stochastic_1998}.
All these methods are based on the physical knowledge or utilize
the pre-understanding of the problem which are expressed in
their own assumptions and fitting parameters. In other words,
all these methods heavily rely on `domain knowledge'.

Machine-learning (ML) approach is based on a different philosophy.
The ML procedure is to develop a machinery which can self-learn
the governing rule or the proper representation of a given problem through the
massive dataset `training'
\cite{rifai_manifold_2011,goodfellow_deep_2016,zhu_handcrafted_2016,lecun_deep_2015}.
Due to the remarkable progress in both hardware and software
engineering, ML technique overwhelms  the state-of-the-art
human-designed algorithms in many different areas
\cite{silver_mastering_2016,lecun_deep_2015}. Recently, it becomes
more and more popular in physics research. ML proves its capability
in many different fields ranging from materials science
\cite{behler_generalized_2007,faber_machine_2016,seko_representation_2017,kolb_discovering_2017,behler_perspective:_2016,artrith_efficient_2017}
and statistical physics
\cite{torlai_learning_2016,huang_accelerated_2017,carrasquilla_machine_2017}
to quantum many-body problems \cite{arsenault_machine_2014,li_pure_2016,wang_discovering_2016,carleo_solving_2017,chng_machine_2017,wang_experimental_2017,zhang_quantum_2017}
and quantum informations \cite{cai_entanglement-based_2015,dunjko_quantum-enhanced_2016,lau_quantum_2017,torlai_neural_2017,tubiana_emergence_2017}.

In this paper, we apply modern ML techniques 
\cite{bengio_greedy_2007,montufar_number_2014} to the long-standing
physics problem of analytic continuation. By using convolutional
neural network (CNN)
\cite{kalchbrenner_convolutional_2014,lecun_deep_2015}
and stochastic gradient descent based optimiser
(\textit{i.e.}, stochastic gradient descent, \textit{Adadelta}, \textit{Adagrad})
\cite{zeiler_adadelta_2012,kingma_adam_2014,duchi_adaptive_2011}, 
we successfully construct the ML kernel which can generate the real frequency 
space spectral function from imaginary Green's function. 
We emphasise that our method does not require any `domain-knowledge' which is a distinctive feature from early-stage ML methods such as statistical learning \cite{arsenault_projected_2017}.
In comparison to the conventional techniques, ML-based algorithm
demonstrates its superiority in terms of accuracy and computation
speed. The spectral weights and peak positions are in better agreement,
and the computation speed is 10$^4$--10$^5$ faster. Further,
ML-based method is more robust against the noise which is inevitably
introduced in Monte-Carlo calculation for example. Our results
show that the domain-knowledge free ML approach can be a new
promising way to solve the long-standing physics problem that has
not been well understood based on the currently available
techniques.

\section{Method}

\subsection{ Description of the problem}

Matsubara frequency Green's function $G(i\omega_{n})$ is analytically
continued to the real frequency $G(\omega)$. For a given $G(i\omega_{n})$,
the spectral function is $A(\omega)=-\frac{1}{\pi}\Im G(\omega+i0^{+})$.
Note that calculating the Green's function for a given spectral function
is straightforward, not ill-conditioned. On the other hand, the spectral
function is obtained by inverting the integral equation
\begin{align}
G(i\omega_{n}) & =\int d\omega\frac{A(\omega)}{i\omega_{n}-\omega}\label{eq:TransFrom}\\
& =\int d\omega{\bf K}(i\omega_{n},\omega){A}(\omega) \label{eq:green_inversion}
\end{align}
where the kernel ${\bf K}(i\omega_{n},\omega)$ has different forms for
different problems. 
This continuation process is an ill-posed problem, and the direct minimizing $\chi^2 =  \sum_{i\omega_{n}}^{N_{{\rm freq}}}||G - {\bf K}A||^{2}$ is hardly feasible due to
the high condition number. The key question is how to deal with intrinsic noises.

\subsection{Description of the machine learning}

Here we note that many techniques to handle this kind of ill-posed 
problems have been actively developed in the ML field of research
for more than last two decades \cite{lecun_gradient-based_1998,vapnik_overview_1999,pillonetto_kernel_2014}.
The early stage ML was basically rule-based, and many details of the problem representation were implemented through handcrafted algorithms.
On the other hand, the modern ML algorithms automatically capture the representations via training, which is often called as `self learning' \cite{lecun_deep_2015,rifai_manifold_2011}.
Since any human knowledge is not directly implemented in the kernel construction, this type of approach is called as domain-knowledge-free ML.
In this modern approach, crucially required are the efficient data representation in high-dimensional
space and the practical algorithm to optimize massive variables in deep
neural networks. In spite of the challenging features of the problems,
modern ML has dramatically surpassed the other state-of-the-art technologies in many areas such as image recognition \cite{krizhevsky_imagenet_2012,farabet_learning_2013,szegedy_going_2015,tompson_joint_2014,kalchbrenner_convolutional_2014}, speech recognition  \cite{mikolov_strategies_2011,hinton_deep_2012,sainath_deep_2013}, language processing \cite{collobert_natural_2011} and translation \cite{cho_learning_2014-1,jean_using_2014,sutskever_sequence_2014}.


In the current study, we adopted `fully connected layer (FCL)' and CNN  \cite{krizhevsky_imagenet_2012,kalchbrenner_convolutional_2014,lecun_deep_2015,carleo_solving_2017},
and try to perform analytic continuation within high-dimensional space.
	The CNN is one of the main players in the high-dimensional data processing for images \cite{ciresan_convolutional_2011,krizhevsky_imagenet_2012,simonyan_very_2014,szegedy_going_2015,he_deep_2016,lecun_deep_2015} and sound/video data  \cite{lecun_deep_2015,masci_stacked_2011,donahue_synthesizing_2018,karpathy_large-scale_2014}.
	We investigated both FCL only and FCL+CNN ML for the analytic continuation problem without using domain-knowledge.
	As a modern domain-knowledge-free ML technique, our approach is well distinguished from the conventional rule-based regression methods \cite{arsenault_projected_2017}.
	It is noted that our neural networks self learn the `rule' or `knowledge' from the massive training with a well prepared extensive data sets.

\subsection{Training}

In order to systematically check the input noise dependence, we
considered several different sets of random noise inputs and examined
the output spectra. 
The Gaussian random noise $N(iw_n)$ is used for our main presentation with the noise strength $\sigma$ (width of Gaussian distribution) varied from 0 to 0.01.
The noised input is then defined as $G(i\omega_{n})^{\rm in}_{\rm noise} = G(i\omega_{n})+N(i\omega_{n})$.
We also considered the other types of noise character.
In particular, the frequency-dependent $\lambda(i\omega_{n})$
has been carefully investigated since it is often the case of 
QMC-DMFT (dynamical mean-field theory) calculations. 
We also considered the uniformly distributed random noise.
We found that the results of the uniform noise ($N(iw_n) \in [-0.05,0.05]$) are comparable with Gaussian noise $\sigma$=0.01.
 While we mainly present the Gaussian random noise, any part of our conclusion is not changed by this choice of noise type.

We constructed the ML-based analytic continuation kernel by
using widely-adopted open-source deep-learning framework, namely,
`\textit{keras}' \cite{chollet2015keras} with `\textit{tensorflow}'
\cite{tensorflow2015-whitepaper} backend. For continuation problem,
the training process is straightforward since
the calculation of $G(i\omega_{n})$ from a given $A(\omega)$ is
not ill-conditioned. Our training sets consist of $\sim$106,000 different combinations of peak numbers, heights and positions.
We generated 18,000 different training data sets for single-peak spectra, 18,000 for double-peak, 20,000 for triple, and 10,000 for each of 4$-$8 peak spectra.
For each set of peak numbers the position, height, and width of the peak are randomly generated in the range of [$-$10, 10], [0.2, 1.0], and [0.3, 1.2], respectively.
 It is straightforward to extend the number of
training sets to an arbitrary number. 
The validation set consists of 10,000 different types of peaks with different random sequences.
For all cases, the normalization condition of $\int A(w) dw=1$ was imposed. It should be noted that, while this particular physical knowledge of normalization is implemented in the training sets, our neural networks is constructed as `domain-knowledge-free' and the kernel should learn the knowledge from the training.

For training, we used `\textit{Adadelta}' \cite{zeiler_adadelta_2012} optimiser. We found that `\textit{stochastic gradient descent (SGD)}'
and even `\textit{RMSprop}' \cite{tieleman_lecture_2012} optimiser
quite often suffer from the `gradient vanishing problem';
\textit{i.e.}, all variables of a neural net are quickly set to zero.
On the other hand, the recently-developed adaptive stochastic variant
optimisers (such as `\textit{Adadelta}', `\textit{Adagrad}'
\cite{duchi_adaptive_2011}, `\textit{Adam}' \cite{kingma_adam_2014},
and `\textit{Adamax}'  \cite{kingma_adam_2014}) produce the reliable
results. We eventually chose `\textit{Adadelta}' as it clearly exhibits
the best performance. For the activation function, we chose a combination
of rectified linear unit (ReLU)
\cite{hahnloser_digital_2000,lecun_deep_2015,ramachandran_searching_2017}
and scaled exponential linear unit (SeLU)\cite{klambauer_self-normalizing_2017}.
It is found that $\sim$8000 epochs are mostly enough for neural network training 
which corresponds to ~16 hours ($\sim$ 7 sec/epoch) at the single desktop PC level (we used one Nvidia 1080 GTX card).

\section{Result and discussion}

\subsection{Fully connected layers}

\begin{figure}[!th]
	\begin{center}
		\includegraphics[width=0.85\linewidth,]{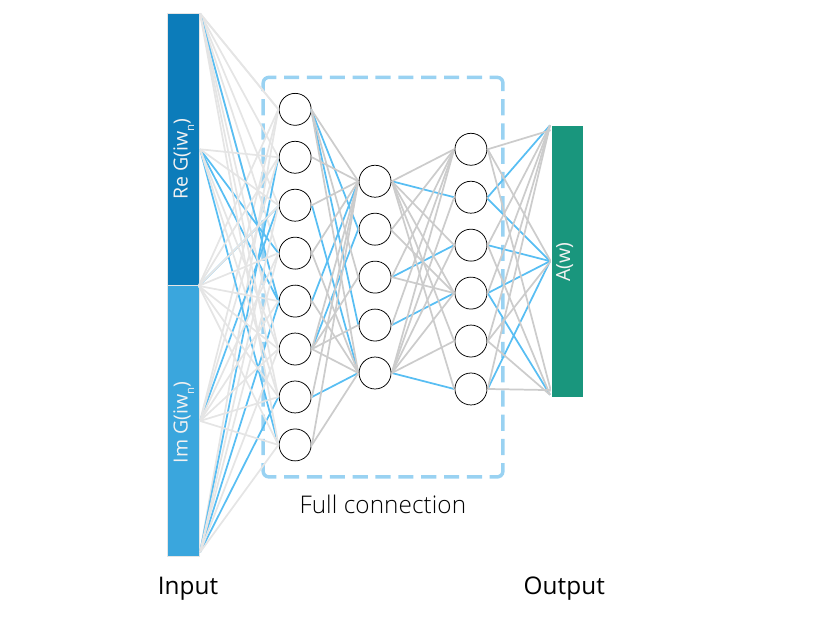} 
	\end{center}
	\caption{ {(Color online) FCL neural network.}
		An illustration of our neural network architecture which
		consists only of FCLs. The input $G(i\omega_{n})$ is an
		array of complex numbers. $\Re[G(i\omega_{n})]$ and
		$\Im[G(i\omega_{n})]$ are arranged as a 1D array to be
		inserted into the neural-net input. The dropout layers
		are located in between all the FCLs to reduce the overfitting
		of neural networks (not shown). The green box represents the
		output layer of $A(\omega)$. The blue neural network lines
		are the schematic representation of activated connections.
		\label{Fig_scheme_nonCNN}
	}
\end{figure}

\label{subsec:dense_only}
As the first step toward ML-based analytic continuation,
we consider the neural network composed of FCLs which may be
regarded as an early-stage ML approach
\cite{frean_upstart_1990,miller_real-time_1990,ripley_neural_1994}.
Roughly, the use of single FCL can be regarded as one 
multiplication process of an inversion matrix to the input Green's
function \cite{kung_kernel_2009}. Having more FCLs thus corresponds
to the increased number of matrix multiplications to represent the inversion.
Practically it is not expected to achieve a notable improvement just by
increasing the number of hidden layers
\cite{muhlenbein_limitations_1990,lecun_gradient-based_1998,dean_large_2012,he_deep_2016}.
After testing many different numbers of hidden layer sets, we indeed
found that the performance is not much enhanced. Thus, in the below,
we focus on the results of three layers (Fig.\ref{Fig_scheme_nonCNN}).

\begin{figure}[t]
	\includegraphics[width=0.99\linewidth]{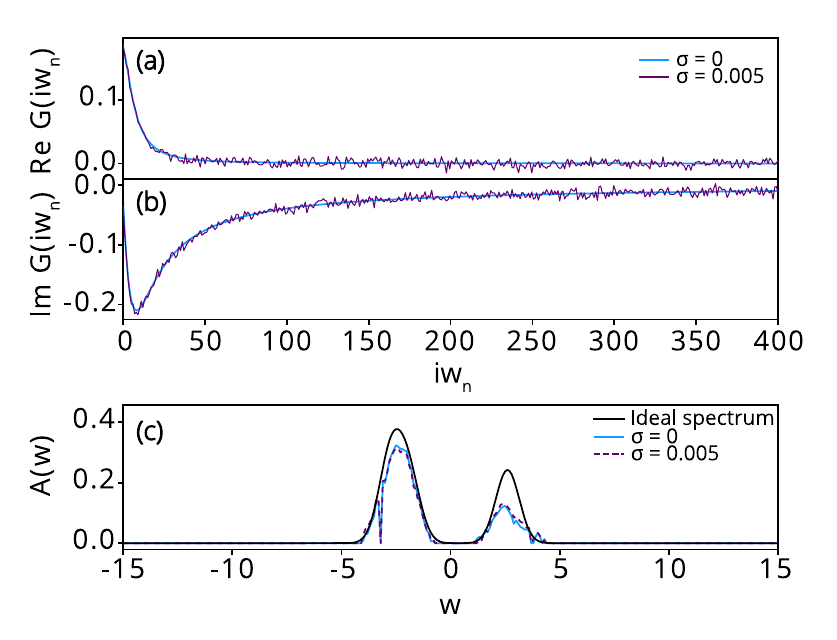} 
	\caption{ {(Color online) Analytic continuation result of FCL.}
		The input Green's functions and the output spectra
		calculated by the FCL neural network kernel (without CNN layer).
		(a, b) $\Re[G(i\omega_{n})]$ and $\Im[G(i\omega_{n})]$
		are presented in (a) and
		(b), respectively. The blue curves are generated
		from the ideal spectrum shown by
		the black line in (c). The purple lines
		show the noised input $G(i\omega_{n})^{\rm{in}}_{\rm{noise}}$
		with $\sigma=0.005$ (see the main text for more details).
		(c) The calculated spectral functions are 
		presented by blue-solid ($\sigma$=0; noise-free) and
		purple-dashed curve ($\sigma$=0.005) in comparison
		to the ideal spectrum (black-solid line).
	}
	\label{Fig_Insulator_nonCNN}
\end{figure}

Figure~\ref{Fig_Insulator_nonCNN} presents the result of analytic
continuation by using FCLs neural network. The black line in
Fig.~\ref{Fig_Insulator_nonCNN}(c) is the spectrum from which
imaginary Green's functions of Fig.~\ref{Fig_Insulator_nonCNN}(a)
and (b) (blue lines) are generated. Therefore, if the continuation
procedure is perfect, the continued spectrum should be identical
with the black line in Fig.~\ref{Fig_Insulator_nonCNN}(c). Note
that the process of obtaining $G(i\omega_{n})$ from $A(\omega)$
is not ill-conditioned. Once $G(i\omega_{n})$ is calculated,
one can perform the continuation and compare the result of
$A(\omega)$ with the original one, namely the ideal spectrum.

The FCL continuation results are shown in
Fig.~\ref{Fig_Insulator_nonCNN}(c). The blue-solid and 
purple-dashed line corresponds to $\sigma$=0 and
$\sigma$=0.005, respectively. It is clearly noted that
the continued spectra are not smooth and significantly
deformed in comparison to the ideal black line. This
result demonstrates the challenging nature of the problem.
At the same time, however, we also note that the overall
shape of spectrum is captured by our FCL neural network
although the unexpected wriggles are found, and they become
worse as the noise level increases. We emphasize that this
level of performance is hardly achievable through the direct
matrix inversion of Eq.~(\ref{eq:green_inversion}). This
promising aspect is largely attributed to the `dropout' and
the regularisation procedure which prevent overfittings
\cite{srivastava_dropout_2014}. In this regard, while not
satisfactory at all, our FCL result shows a possibility of
neural network approach for the analytic continuation.

\begin{figure}[t]
	\includegraphics[width=0.90\linewidth]{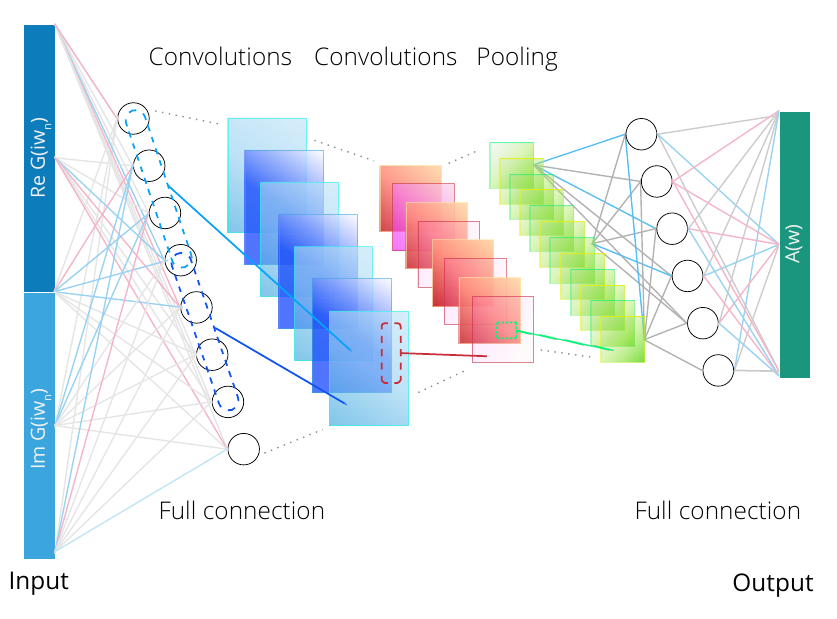} 
	\caption{ {(Color online) CNN neural network.} An illustration of the
		neural network architecture which consists of both FCLs
		and CNN. Two convolution layers (blue and red squares)
		and one pooling layer (green squares) are placed in 
		between the FCLs. Coloured neural network lines are the
		schematic representations of activated connections.
	}
	\label{Fig_scheme_CNN}
\end{figure}

\subsection{Convolutional neural network}

Many techniques have been suggested to overcome the deficiency of FCL.
One key idea is to identify the essential features of a problem 
and to reconstruct them in a higher dimensional space
\cite{montufar_number_2014,pillonetto_kernel_2014}.
Principal component analysis (PCA)
\cite{shlens_tutorial_2014,hall_properties_2006} is an example which
proved to be powerful for data compression and dimensionality reduction.
Unfortunately, however, PCA can only be used in rank 1 (vector) and
rank 2 (matrix) for most of the cases. While some techniques for tensor
PCA have been proposed, they seem to need further developments
\cite{richard_statistical_2014,inoue_generalized_2016,zhang_tensor_2018,li_convolutional_2015}.
A typical fundamental limitation of PCA is that each principal component
is given by a linear combination of original variables whereas non-linearity
is essential for ill-posed problems \cite{archambeau_sparse_2009}. In this
regard, CNN is a useful advanced technique leading the modern machine-learning era
\cite{ciresan_convolutional_2011,krizhevsky_imagenet_2012,simonyan_very_2014,szegedy_going_2015,he_deep_2016,lecun_deep_2015}.
The performance of CNN image processing surpasses the human-designed
algorithms based on `domain knowledge'
\cite{ciresan_convolutional_2011,krizhevsky_imagenet_2012}. Due to its
outstanding feature selection in tensor space, CNN is widely adopted
by high-dimensional noise filters for auto-encoder and sound/video data \cite{masci_stacked_2011,donahue_synthesizing_2018,karpathy_large-scale_2014}.

In analytic continuation, input/output data are represented by a certain
set of numbers. Thus it can be regarded as an inverse problem that has
to be performed within a dimension corresponding to those numbers. With
this observation, we applied CNN technique to the long-standing ill-posed
problem of analytic continuation. Figure~\ref{Fig_scheme_CNN} shows our
neural network structure. We aim to create a minimal model with the
smallest possible number of layers. Thus our neural network is designed
to contain CNN layers in between two FCLs since we learned in the above
that three FCLs could capture the basic features of spectra. While it is
conventional to have CNN layers just next to the input layer in the image
processing (\textit{e.g.}, AlexNet \cite{krizhevsky_imagenet_2012}, VGG
\cite{simonyan_very_2014}, GoogleNet \cite{szegedy_going_2015}, and ResNet
\cite{he_deep_2016}), we take a different strategy of inserting the CNN
layer after the matrix operation through FCL. It is because the full
information of input Green's function needs to be utilized in our problem.
The total number of parameters in our neural network is $\sim$600,000
and $\sim$500,000 for including and excluding CNN, respectively. It is
noted that the network size is not much increased by having CNN layers.
We have adopted modern optimization algorithm, namely `\textit{Adadelta}'
\cite{zeiler_adadelta_2012}, to optimize this large number of neural
networks parameters.

\begin{figure}[t]
	\includegraphics[width=0.99\linewidth]{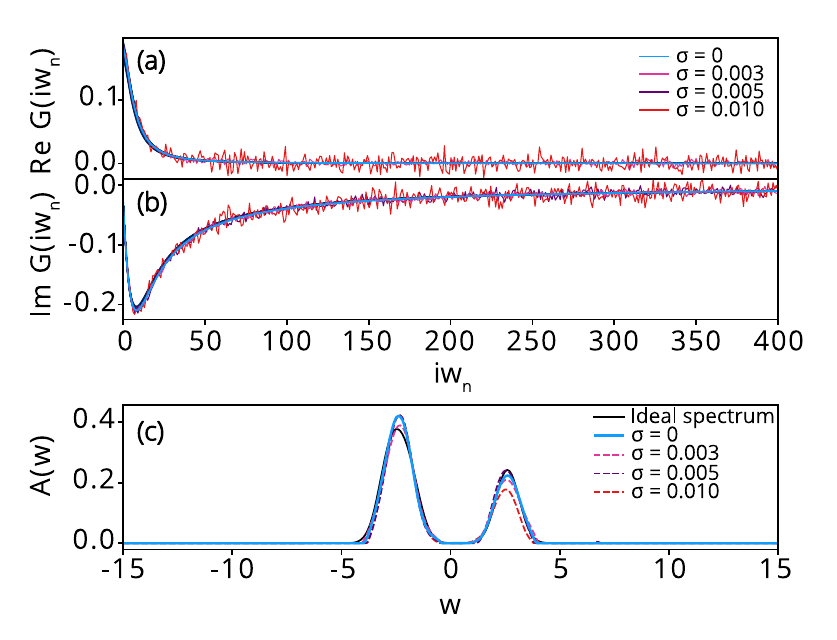} 
	\caption{{(Color online) Analytic continuation result of CNN.}
		(a, b) $\Re[G(i\omega_{n})]$ and $\Im[G(i\omega_{n})]$
		are presented in (a) and (b), respectively. The light-blue
		curves are generated from the ideal spectrum shown by the black
		line in (c). 
		Five different noise levels are presented in light-blue ($\sigma$=0),  magenta ($\sigma$=0.003), purple ($\sigma$=0.005) and red lines 
		($\sigma$=0.01). (c) Analytic continuation result of CNN-ML
		kernel for $\sigma$=0 (noise-free; light-blue-solid),
		 $\sigma$=0.003 (magenta-dashed), $\sigma$=0.005 (purple-dashed) and $\sigma$=0.01
		(red-dashed) along with the ideal spectrum (black-solid line).
	}
	\label{Fig_Insulator1_CNN}
\end{figure}

Figure~\ref{Fig_Insulator1_CNN} shows the continuation result of
using CNN. The model spectrum (black line in
Fig.~\ref{Fig_Insulator1_CNN}(c)) is designed to mimic a
Mott-Hubbard insulator consisted of two distinct Hubbard bands with
different peak heights.
The outstanding performance of CNN can clearly be seen from that the
continuation results are significantly improved in comparison to
the FCL-only data in Fig.~\ref{Fig_Insulator_nonCNN}. The overall
shape, peak positions and relative peak heights are well reproduced
without any undesirable wriggle. Importantly, the reproducibility
remains quite robust against the noise even if the deviation from
the ideal spectrum (black) becomes noticeable as the noise level increases 
(from light-blue-solid lines to red-dashed).
We also checked that the reconstructed ${\tilde{G}}(i\omega_n)=({\rm\bf{K}}A)(i\omega_n)$ is consistent with  $G(i\omega_{n})$ within the noise level (not shown). 
For example, the calculated $ {\tilde {\chi}}$ = 0.0044 for the case of $\sigma=0.003$  
where $ \tilde{\chi} \equiv (\chi^2/N_{\rm freq})^{-1/2}$.

\begin{figure}[t]
	\centering
	\includegraphics[width=0.99\linewidth]{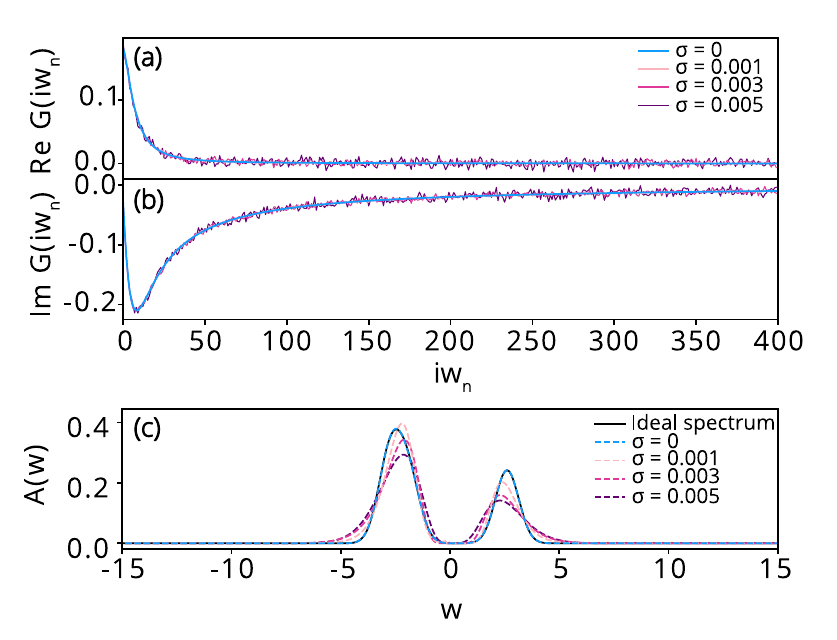} 
	\caption{{(Color online) Analytic continuation result of MEM.} 
		(a, b) Input Green's function of 	
		$\Re[G(i\omega_{n})]$ and $\Im[G(i\omega_{n})]$
		are presented in (a) and
		(b), respectively. 
		The light blue curves are generated
		from the ideal spectrum shown by
		the black line in (c).
	Four different noise levels
		are presented in light-blue ($\sigma$=0), 
		light-pink ($\sigma$=0.001), magenta ($\sigma$=0.003) and purple lines ($\sigma$=0.005).
		(c) Analytic continuation result of conventional MEM
		for$\sigma$=0 (noise-free; light-blue-solid),
		$\sigma$=0.001 (light-pink-dashed), $\sigma$=0.003 (magenta-dashed) and $\sigma$=0.005 (purple-dashed)  along with
		the ideal spectrum (black-solid line).
	}
	\label{Fig_Insulator_Mem}
\end{figure}

The robustness against the input noise is a crucially required feature 
for the reliable analytic continuation since the noise is unavoidably
present in stochastic approaches. As shown in Fig.~\ref{Fig_Insulator1_CNN}(c),
the overall features and the detailed shapes of the spectrum are well
maintained even for the case of significant noise levels.
This result shows the powerfulness of ML-based analytic continuation kernel.

The performance of ML kernel is further demonstrated by the comparison to
the conventional continuation technique, namely MEM.
The details of our MEM algorithms can be found in Ref.~ \onlinecite{bergeron_algorithms_2016,sim_maximum_2018}.
Fig.~\ref{Fig_Insulator_Mem}
shows the result of MEM which is one of the most widely-used methods for
analytic continuation
\cite{haule_dynamical_2010,bergeron_algorithms_2016,jarrell_bayesian_1996,gunnarsson_analytical_2010}. 
It is clearly noted that, even at a significantly lower noise
level, the MEM result is markedly deviated from the ideal spectrum in
terms of peak position and height. It is in a sharp contrast to the
ML-based result of Fig.~\ref{Fig_Insulator1_CNN} in which the spectrum
is well preserved even at $\sigma=0.01$.

\begin{figure}[t]
	\includegraphics[width=0.99\linewidth]{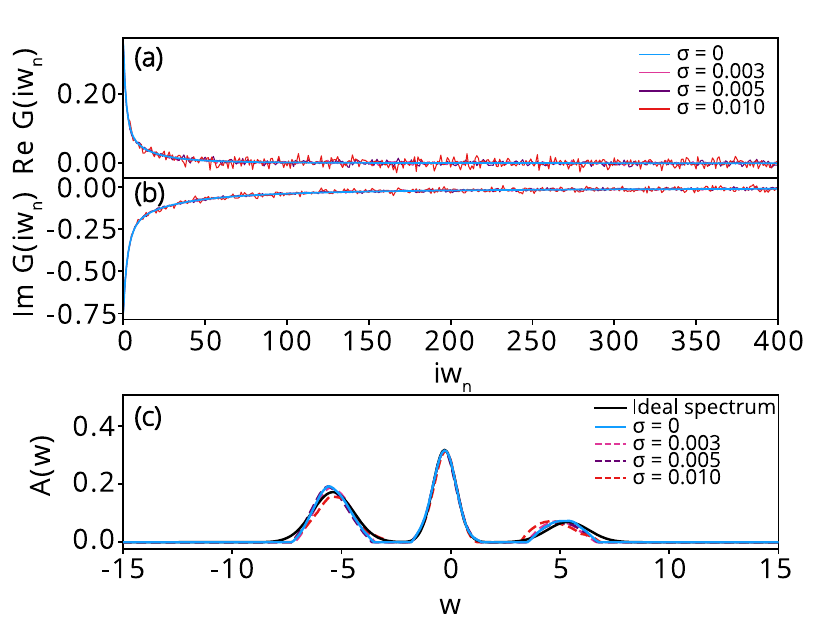} 
	\caption{{(Color online) Analytic continuation result of CNN for metallic spectrum.}
		(a, b) Input Green's function of 	
		$\Re[G(i\omega_{n})]$ and $\Im[G(i\omega_{n})]$
		are presented in (a) and
		(b), respectively. 
		The light-blue curves are generated from the
		ideal spectrum shown by the black line in (c).	
		Five different noise
		levels are presented in light-blue ($\sigma$=0), magenta ($\sigma$=0.003), purple ($\sigma$=0.005) and red lines 
		($\sigma$=0.01). (c) Analytic continuation
		result of CNN-ML kernel for$\sigma$=0 (noise-free; light-blue-solid), $\sigma$=0.003 (magenta-dashed), $\sigma$=0.005 (purple-dashed) and $\sigma$=0.01 (red-dashed)
		along with the ideal spectrum (black-solid line).
	}
	\label{Fig_Metal1_CNN}
\end{figure}

Figure~\ref{Fig_Metal1_CNN} shows the result of ML kernel for metallic
spectrum having coherent as well as incoherent peaks. Once again, our
machine-learning kernel well reproduces the original spectrum. The
robustness against noise is also excellent as in the insulating case.
In particular, the coherent peak is considerably well reproduced
while the incoherent states are moderately affected by the noises.
It is a good indication for predicting the phase from a given
Green's function.

\begin{figure}[t]
	\includegraphics[width=0.9\linewidth]{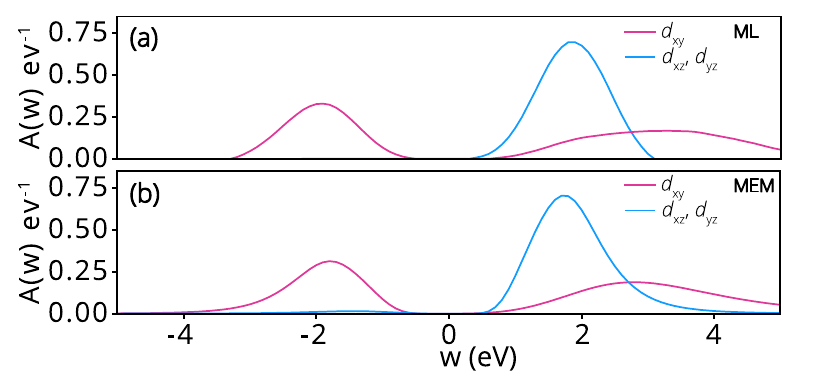} 
	\caption{
	{(Color online) 
	The calculated DFT+DMFT spectral function of monolayer SrVO$_3$ by (a) CNN and (b) MEM.
	The magenta and blue colors refer to the $d_{xy}$ and $d_{yz,xz}$ orbital character, respectively.
	Comparison between ML and MEM results for DFT+DMFT spectral function of the SrVO$_3$ monolayer}
	The input Green's function is obtained from DFT+DMFT with CT-QMC solver.
	Analytic continuation for the spectral function was performed using (a) ML-CNN and (b) MEM.}
	\label{Fig_SVO_MEM_ML}
\end{figure}

As the last example, we present the results of a real material, namely SrVO$_3$ monolayer.
While the bulk SrVO$_3$ is a correlated metal,
it becomes an insulator in the monolayer limit \cite{PhysRevLett.104.147601,schuler_charge_2018}.
The Green's function is obtained from DFT+DMFT calculation \cite{LichtensteinKatsnelson_Initio1998,GeorgesKotliar_Hubbard1992}
combined with the 
hybridization expansion continuous-time quantum Monte Carlo algorithm 
\cite{werner_continuous-time_2006,werner_hybridization_2006,haule_quantum_2007, comp_details}.
The spectra obtained by our ML kernel is reasonably well compared with those of MEM; see Fig. \ref{Fig_SVO_MEM_ML}.
While the peak widths are slightly narrower in MEM, 
it should be noted
that the direct use of MEM for Green’s function tends to broaden the spectra \cite{wang_antiferromagnetism_2009}.


Of particular interest is the performance of our ML continuation kernel for the cases that were not included in the training sets. Although the systematic investigation of the training-set dependence is not the main interest of the current study, we obtained some meaningful results.
In terms of peak numbers, the quality of continuation gets gradually worse as the number of peaks goes out of the training range. 
However, its performance is still quite decent (we tried up to the 25-peak case) and at least comparable with that of MEM. The similar feature is found for the peak width.
For the peak of width=0.1 and 0.2, the ML kernel produces the spectrum whose width is 0.22 and 0.24, respectively.
Considering that the conventional continuation techniques suffer from the same problems in describing sharp peaks, we concluded that the ML exhibits reasonably good performance.
Finally, we emphasize the efficiency of ML-based analytic continuation.
Once the ML kernel is well trained, the continuation process can be
performed at a speed of $\sim$10,000 Green's functions/sec, which is
at least $10^4$--$10^5$ faster than the conventional MEM.

\section{Conclusion}

Modern ML technique proves its usefulness for a long-standing physics
problem of analytic continuation. Its superiority over the
conventional technique is demonstrated in terms of the accuracy, speed,
and the robustness against noise. For both insulating and metallic
spectrum, our CNN-based ML kernel gives the better agreement with the
ideal spectrum in terms of peak position and height. Up to the
high level of random noises, at which MEM fails to produce
the reliable results, ML technique retains its accuracy. In terms
of computation speed, the trained kernel is 10$^4$--10$^5$ times
faster than the conventional method. 
Our result suggests that `domain-knowledge' free ML can be used as an alternative tool for the physics problems where the conventional methods have been struggling. We also note the possibility of certain types of hybrid methods in which a part of physics intuitions would be combined with ML approach.

\section{Acknowledgments}

This work was supported by Basic Science Research Program through
the National Research Foundation of Korea (NRF) funded by the Ministry
of Education (2017R1D1A1B03032082) and
Creative Materials Discovery Program through the NRF funded by Ministry of Science and
ICT (2018M3D1A1059001).


\bibliography{ML_analyticCont}

\end{document}